\documentclass[11pt]{article}             
\usepackage{osid}                         %
\usepackage{amsmath,amssymb,bbm}
\usepackage[dvips]{epsfig}

\title{Decoherence-Free Subspaces and Subsystems\\
for a Collectively Depolarizing Bosonic Channel}

\author{Jonathan L. Ball and Konrad Banaszek\\{\footnotesize\it
Clarendon Laboratory, University of Oxford\\Oxford OX1 3PU, United Kingdom\\
e-mail: jonathan.ball@keble.oxford.ac.uk, k.banaszek@qubit.org}}

\begin{document}

\maketitle
\begin{abstract}
We discuss the structure of decoherence-free subsystems for a bosonic
channel affected by collective depolarization.
A single use of the channel is defined as a transmission
of a pair of bosonic modes. Collective
depolarization consists in a random linear U(2) transformation of the respective
mode operators, which is assumed to be identical for
$N$ consecutive uses of the channel. 
We derive a recursion formula that characterizes the dimensionality of
available decoherence-free subsystems in such a setting.
\end{abstract}

\section{Introduction}

Decoherence induced by uncontrolled interactions with the environment
is a major obstacle when implementing protocols for quantum information
processing in real physical systems. The effects of such interactions can however
be reduced by using particular quantum states that are robust against specific decoherence mechanisms. An interesting and physically relevant scenario of decoherence is when the physical system exhibits certain symmetries in interactions with the environment. Such symmetries imply the existence of whole subspaces that remain completely unaffected by decoherence and can therefore be used for faultless quantum information processing. This observation, which can be made for a number of interaction schemes from various perspectives \cite{PalmSuomPRS96,DuanGuoPRL97,ZanaRasePRL97,LidaChuaPRL98,KnilLaflPRL00}, has led to the development of a general theory of decoherence-free subspaces and subsystems, reviewed recently in \cite{DFSReview}.

The theory of decoherence-free subspaces and subsystems applies to a variety of physical systems. One such system is light travelling through
an optical fiber, which usually exhibits random birefringence caused by fluctuating environmental conditions, such as temperature and mechanical strain.
Mathematically, this system can be modelled as a bosonic communication channel. A single use of this channel is defined as a transmission of two bosonic modes corresponding to orthogonal polarizations. Birefringence is described as a random U(2) transformation between the operators of these two orthogonally polarized modes. In a realistic situation this transformation remains practically constant over many consecutive uses of the channel, for time intervals comparable at least with the round trip time in a fiber \cite{GisiRiboRMP02}. The presence of such a symmetry leads to the existence of non-trivial decoherence-free subsystems. Their structure has been discussed in our recent paper \cite{BallBanaXXX04}. In this contribution, we will concentrate on the mathematical details of the derivation briefly sketched in the previous paper, and will illustrate the presentation with diagrams that provide additional insights into the calculations. 

The decoherence-free subsystems for a collectively depolarizing bosonic channel have recently attracted considerable attention in the context of quantum reference frames
and various scenarios of quantum and classical communication over depolarizing channels \cite{BartRudoPRL03}. This has produced theoretical proposals for robust quantum key distribution \cite{WaltAbourPRL03,BoilGottPRL04,BoilLaflPRL04}, alignment-free tests of Bell's inequalities \cite{CabePRL03}, and entanglement-enhanced classical communication \cite{BallDragPRA04}, with some of these ideas implemented in proof-of-principle quantum optical experiments \cite{KwiaBergSCI00,BourEiblPRL04,BanaDragPRL04}.
{}From the mathematical point of view, an interesting aspect of the collectively depolarizing bosonic channel is that in contrast to particle-based scenarios in which every elementary system (i.e.\ a particle) has a finite number of states, the elementary system in our case (i.e.\ a pair of bosonic modes) is infinitely dimensional and its transformation under the interaction with the environment is described by a reducible representation of the group U(2). 

This paper is organized as follows. We start with a discussion of the polarization transformation for a single use of the bosonic channel in Sec.~2. We then analyze in Sec.~3 the emergence of decoherence-free subsystems for multiple uses of the channel. In Sec.~4 we derive a recursion formula for dimensions of
decoherence-free subsystems. Finally, Sec.~5 concludes the paper.

\section{Polarization transformation}
\label{Sec:PolarizationTransformation}

We assume that the quantum system transmitted in a single use of the channel is composed of a pair of bosonic modes corresponding to orthogonal polarizations, with the respective creation operators labelled by $\hat{a}_{H}^\dagger$ and $\hat{a}_{V}^\dagger$. Random depolarization consists in a linear unitary transformation of the modes given by:
\begin{equation}
\label{Eq:PolarizationTransformation}
\left( \begin{array}{c} \hat{a}_{H}^\dagger \\
\hat{a}_{V}^\dagger \end{array} \right)
\mapsto
{\bf\Omega}
\left( \begin{array}{c} \hat{a}_{H}^\dagger \\
\hat{a}_{V}^\dagger \end{array} \right)
\end{equation}
where the matrix
\begin{equation}
{\bf\Omega} = 
\left( \begin{array}{cc} \Omega_{HH} & \Omega_{HV} \\
\Omega_{VH} & \Omega_{VV}
\end{array} \right)
\end{equation}
is an arbitrary element of the group U(2). It will be convenient to decompose ${\bf\Omega}$
into a product of a phase factor $e^{-i\alpha} \in \mbox{U(1)}$ and a special unitary
matrix ${\bf\Omega}' \in \mbox{SU(2)}$:
\begin{equation}
\label{Eq:OmegaDecomposition}
{\bf\Omega} = e^{-i\alpha} {\bf\Omega}'.
\end{equation} 
This decomposition is ambiguous, as both the factors can be multiplied by $-1$. However, the final results obtained with this decomposition will be free from this ambiguity, which makes the specific choice of the decomposition irrelevant. It is also worthwhile to note that in contrast to unitary transformations on the Hilbert space of quantum states, for which the overall phase factor is not physical, in the present case the overall phase factor $e^{-i\alpha}$ is physically meaningful, as it describes the phase of the fields which can in principle be measured with an external phase reference. 

The two-mode Hilbert space ${\cal H}$ describing our system has a convenient orthonormal basis in the form of Fock states defined in general by
\begin{equation}
|m_H n_V\rangle=
\frac{(\hat{a}_{H}^\dagger)^{m}(\hat{a}_{V}^\dagger)^{n}}{\sqrt{m!n!}}
|{\textrm{vac}}\rangle, 
\qquad
m,n = 0,1,2,\ldots,
\end{equation}
where $|{\textrm{vac}}\rangle$ is the vacuum state of the system.
The polarization transformation defined in Eq.~(\ref{Eq:PolarizationTransformation}) does not change the total number of field excitations. It is therefore convenient to decompose the  Hilbert space ${\cal H}$ into a direct sum
\begin{equation}
\label{Eq:H=sumH(l)}
{\cal H} = \bigoplus_{l=0}^{\infty} {\cal H}^{(l)}
\end{equation}
of finite-dimensional
subspaces ${\cal H}^{(l)}$ that contain exactly $l$ excitations in both the modes. The subspace ${\cal H}^{(l)}$ has dimension $l+1$ and it is spanned by Fock states of the form:
\begin{equation}
\label{Eq:H(l)=Span}
{\cal H}^{(l)} = \mbox{Span} \{ |m_{H} (l-m)_{V}\rangle \; | \; m=0,1,\ldots, l \}.
\end{equation}
Under the polarization transformation given in Eq.~(\ref{Eq:PolarizationTransformation}), the state $|m_{H} (l-m)_{V}\rangle$ is transformed according to:
\begin{eqnarray}
\lefteqn{|m_{H} (l-m)_{V}\rangle = \frac{(\hat{a}_{H}^\dagger)^{m}(\hat{a}_{V}^\dagger)^{l-m}}{\sqrt{m!(l-m)!}}
|{\textrm{vac}}\rangle}
& & \nonumber \\
\label{Eq:ml-m}
& \mapsto &
\frac{(\Omega_{HH}\hat{a}_{H}^\dagger+\Omega_{HV}\hat{a}_{V}^\dagger)^{m}
(\Omega_{VH}\hat{a}_{H}^\dagger+\Omega_{VV}\hat{a}_{V}^\dagger)^{l-m}}{\sqrt{m!(l-m)!}}
|{\textrm{vac}}\rangle 
\end{eqnarray}
If we now insert the decomposition of ${\bf\Omega}$ given in Eq.~(\ref{Eq:OmegaDecomposition}), this will produce an overall factor $e^{-il\alpha}$
times the same expression as in the second line of Eq.~(\ref{Eq:ml-m}), but with the elements of ${\bf\Omega}$ replaced by those of ${\bf\Omega}'$. It is easy to recognize in this expression the standard construction of irreducible representations of the group SU(2) using monomials \cite{Cornwell,Greiner}. This yields the formula:
\begin{equation}
|m_{H} (l-m)_{V}\rangle \mapsto e^{-il\alpha} \sum_{n=0}^{l}
D^{l/2}_{mn}({\bf\Omega}') |n_{H} (l-n)_{V}\rangle
\end{equation}
where $D^{l/2}_{mn}({\bf\Omega}')$ are the elements of $(l+1)\times(l+1)$ matrices 
$\hat{D}^{l/2}({\bf\Omega}')$ which form the irreducible $(l+1)$-dimensional representation
of the group SU(2). Thus the unitary transformation $\hat{U}({\bf\Omega})$ of an arbitrary quantum state in the Hilbert space ${\cal H}$ induced by the map given in Eq.~(\ref{Eq:PolarizationTransformation}) has the form:
\begin{equation}
\label{Eq:UOmega}
\hat{U}({\bf\Omega}) = \bigoplus_{l=0}^{\infty} e^{-il\alpha} \hat{D}^{l/2}({\bf\Omega}').
\end{equation}
Let us note that although the decomposition ${\bf\Omega}=e^{-i\alpha}{\bf\Omega}'$ is defined up to $-1$ multiplying both the factors, the product $e^{-il\alpha} \hat{D}^{l/2}({\bf\Omega}')$ does not depend on the specific choice of the decomposition.
This follows from the fact that the elements of the matrix $\hat{D}^{l/2}({\bf\Omega}')$ are given by monomials of degree $l$ constructed from the elements of ${\bf\Omega}'$. Therefore each one of the substitutions $e^{-i\alpha} \rightarrow
- e^{-i\alpha}$ and ${\bf\Omega}' \rightarrow - {\bf\Omega}'$
will produce a factor $(-1)^{l}$, one multiplying $e^{-il\alpha}$ and another one multiplying $\hat{D}^{l/2}({\bf\Omega}')$, which will cancel each other.  Consequently, the right hand side of Eq.~(\ref{Eq:UOmega}) is defined unambiguously as a function of ${\bf \Omega} \in \mbox{U(2)}$.

\section{Collective depolarization}
\label{Sec:CollectiveDepolatization}

We will now consider the scenario in which the polarization transformation ${\bf\Omega}$ is constant across $N$ uses of the channel. The entire Hilbert space in this case is given by an $N$-fold tensor product ${\cal H}^{\otimes N}$ of the two-mode space ${\cal H}$ analyzed in the previous section. The action of the collectively depolarizing channel on an 
arbitrary input quantum state $|\psi\rangle \in {\cal H}^{\otimes N}$ is given by:
\begin{equation}
|\psi\rangle \mapsto 
[\hat{U}({\bf\Omega})]^{\otimes N} |\psi\rangle.
\end{equation}
Given the decomposition of $\hat{U}({\bf\Omega})$ derived
in Eq.~(\ref{Eq:UOmega}), we can rewrite the tensor product $[\hat{U}({\bf\Omega})]^{\otimes N}$ as:
\begin{eqnarray}
[\hat{U}({\bf\Omega})]^{\otimes N}
& = & \bigoplus_{l_1,\ldots,l_N \ge 0}
e^{-i(l_1+\ldots+l_N)\alpha} 
\hat{D}^{l_1/2}({\bf\Omega}') \otimes 
\ldots \otimes \hat{D}^{l_N/2}({\bf\Omega}')
\nonumber \\
\label{Eq:UotimesN}
& = & \bigoplus_{L=0}^{\infty} e^{-iL\alpha}
\bigoplus_{\substack{l_1,\ldots,l_N \ge 0\\
l_1+\ldots+l_N=L}}
\hat{D}^{l_1/2}({\bf\Omega}') \otimes 
\ldots \otimes \hat{D}^{l_N/2}({\bf\Omega}') 
\end{eqnarray}
We see that the overall phase factor of the transformation ${\bf\Omega}$ enters
the expression only with the total number of excitations $L=l_1+\ldots+l_N$ contained
in all the $N$ uses of the channel. The $N$-fold tensor product of the SU(2)
representations $\hat{D}^{l_1/2}({\bf\Omega}') \otimes 
\ldots \otimes \hat{D}^{l_N/2}({\bf\Omega}')$ can in general be
decomposed into direct sums by the iterative application of the formula
\cite{BrinkSatchler}:
\begin{equation}
\label{Eq:Dj1Dj2}
\hat{D}^{j_1}({\bf\Omega}') \otimes
\hat{D}^{j_2}({\bf\Omega}') 
=
\hat{D}^{|j_1-j_2|}({\bf\Omega}') \oplus \hat{D}^{|j_1-j_2|+1}({\bf\Omega}') 
\oplus \ldots \oplus \hat{D}^{j_1+j_2}({\bf\Omega}').
\end{equation}
This formula allows one to convert all the tensor products into direct sums
of irreducible representations, with any of the representations allowed to appear
a number of times.
Therefore we anticipate that the inner sum in Eq.~(\ref{Eq:UotimesN})
can be represented in the form:
\begin{equation}
\label{Eq:DecompositionforfixedL}
\bigoplus_{\substack{l_1,\ldots,l_N \ge 0 \\
l_1+\ldots+l_N=L}}
\hat{D}^{l_1/2}({\bf\Omega}') \otimes 
\ldots \otimes \hat{D}^{l_N/2}({\bf\Omega}')
= \bigoplus_{j} \hat{\mathbbm{1}}_{K^{j}_{NL}} \otimes \hat{D}^{j}({\bf\Omega}')
\end{equation}
where $\hat{\mathbbm{1}}_K$ is the identity operator in a $K$-dimensional space
$\mathbb{C}^{K}$ and
the integer $K^{j}_{NL}$ tells us how many times the representation $j$ occurs
in the decomposition of the $N$-fold product $[\hat{U}({\bf\Omega})]^{\otimes N}$
in the sector of the entire Hilbert space ${\cal H}^{\otimes N}$ that contains
exactly $L$ excitations. The multipliticies $K^{j}_{NL}$ will define the capability
of the system to protect quantum coherence against depolarization. By considering
the parity of the indices $l_1,\ldots l_N$ when applying Eq.~(\ref{Eq:Dj1Dj2})
to multiple tensor products $\hat{D}^{l_1/2}({\bf\Omega}') \otimes 
\ldots \otimes \hat{D}^{l_N/2}({\bf\Omega}')$,
it is easy to observe that the representations which will appear in the decomposition
in Eq.~(\ref{Eq:DecompositionforfixedL}) will be indexed with $j=0,1,\ldots, L/2$ for
even $L$ and with $j=1/2,3/2,\ldots,L/2$ for odd $L$. Assuming for the time being
that the multiplicities $K^{j}_{NL}$ are known, we thus
arrive at the following decomposition
of the polarization transformation in the entire Hilbert space ${\cal H}^{\otimes N}$:
\begin{equation}
[\hat{U}({\bf\Omega})]^{\otimes N}
=
\bigoplus_{L=0}^{\infty} e^{-iL\alpha}
\bigoplus_{j=L/2-\lfloor L/2 \rfloor}^{L/2} \hat{\mathbbm{1}}_{K^{j}_{NL}} \otimes
\hat{D}^{j}({\bf \Omega}').
\end{equation}

The above formula suggests the decomposition of the Hilbert space ${\cal H}^{\otimes N}$ into sectors with a fixed number of excitations ${\cal H}_{NL}$:
\begin{equation}
{\cal H}^{\otimes N} 
=
\bigoplus_{L=0}^{\infty}
{\cal H}_{NL}
\end{equation}
which in turn can be represented as isomorphic with the following structure:
\begin{equation}
\label{Eq:HNLCC}
{\cal H}_{NL} \cong
\bigoplus_{j=L/2-\lfloor L/2 \rfloor}^{L/2}
\mathbb{C}^{K^{j}_{NL}} \otimes \mathbb{C}^{2j+1}.
\end{equation}
The components of the state vector belonging to separate subspaces ${\cal H}_{NL}$
are multiplied by different phase factors $e^{-iL\alpha}$. If the parameters
of the transformation ${\bf\Omega}$ are unknown, this implies the loss of quantum
coherence between different sectors ${\cal H}_{NL}$. Within each sector, however,
we have a number of subspaces for which the action of the collective depolarization
operator
is given by the identity $\hat{\mathbbm{1}}_{K^{j}_{NL}}$. This means
that a quantum state which is encoded into a subspace isomorphic to one of the
subspaces $\mathbb{C}^{K^{j}_{NL}}$ in the structure defined in Eq.~(\ref{Eq:HNLCC})
remains intact even if the transformation ${\bf\Omega}$ is completely unknown.
We note that the coherence between the subspaces $\mathbb{C}^{K^{j}_{NL}}$ with
different values of $j$ is destroyed as they are coupled to subsystems
$\mathbb{C}^{2j+1}$ on which collective depolarization takes different forms.
Consequently, the capacity of the system to protect quantum coherence is
defined by the multiplicities ${K^{j}_{NL}}$, or to be precise, by
the highest value over the permitted range of $j$.

\section{Recursion formula}
\label{Sec:RecursionFormula}

We will now demonstrate that the multiplicities $K^{j}_{NL}$ are related
via a simple recursion formula. As we have seen in the preceding section, 
when the total number $L$ of excitations is fixed, the
overall phase of the polarization transformation ${\bf\Omega}$ is irrelevant.
It is therefore sufficient to restrict ourselves to ${\bf\Omega} \in \mbox{SU(2)}$
and consider the decomposition of $[\hat{U}({\bf\Omega})]^{\otimes N}$ in
the sector ${\cal H}_{NL}$ in the form:
\begin{equation}
\label{Eq:UOmegaHNLExpansion}
\left. [\hat{U}({\bf\Omega})]^{\otimes N} \right|_{{\cal H}_{NL}}
=
\bigoplus_{j=L/2-\lfloor L/2 \rfloor}^{L/2} \hat{\mathbbm{1}}_{K^{j}_{NL}} \otimes
\hat{D}^{j}({\bf \Omega}).
\end{equation}
The Hilbert space corresponding to $N$ uses of the channel has an obvious decomposition
as ${\cal H}^{\otimes N} = {\cal H}^{\otimes (N-1)} \otimes {\cal H}$. If we now
consider the sector of ${\cal H}^{\otimes N}$ containing exactly $L$ excitations, it
can be constructed from the subspaces of ${\cal H}^{\otimes (N-1)}$ and ${\cal H}$
according to:
\begin{equation}
{\cal H}_{NL} = \bigoplus_{L'=0}^{L} {\cal H}_{N-1,L'}
 \otimes {\cal H}^{(L-L')}
\end{equation}
where ${\cal H}^{(L-L')}$ is a subspace of the two-mode Hilbert space introduced in Eq.~(\ref{Eq:H=sumH(l)}) that contains exactly $L-L'$ excitations of the field. 
This construction implies that the operator
$\left. [\hat{U}({\bf\Omega})]^{\otimes N} \right|_{{\cal H}_{NL}}$ can be represented
as:
\begin{equation}
\left. [\hat{U}({\bf\Omega})]^{\otimes N} \right|_{{\cal H}_{NL}}
 = 
\bigoplus_{L'=0}^{L}
\left.[\hat{U}({\bf\Omega})]^{\otimes (N-1)} \right|_{{\cal H}_{N-1,L'}}
\otimes
\hat{D}^{(L-L')/2}({\bf \Omega})
\end{equation}
We can now insert the decomposition of 
$\left.[\hat{U}({\bf\Omega})]^{\otimes (N-1)} \right|_{{\cal H}_{N-1,L'}}$
using an expansion analogous 
to Eq.~(\ref{Eq:UOmegaHNLExpansion}), assuming that the respective
coefficients $K^{j'}_{N-1,L'}$ are known. This yields:
\begin{equation}
\label{Eq:LostTheTrack}
\left. [\hat{U}({\bf\Omega})]^{\otimes N} \right|_{{\cal H}_{NL}}
=
\bigoplus_{L'=0}^{L}
\left(
\bigoplus_{j'=L'/2-\lfloor L'/2 \rfloor}^{L'/2} \hat{\mathbbm{1}}_{K^{j'}_{N-1,L'}} \otimes
\hat{D}^{j'}({\bf \Omega})
\right)
\otimes
\hat{D}^{(L-L')/2} ({\bf\Omega}) 
\end{equation}
The tensor product 
$\hat{D}^{j'}({\bf \Omega}) \otimes \hat{D}^{(L-L')/2} ({\bf\Omega})$ appearing in the above expression can be decomposed using Eq.~(\ref{Eq:Dj1Dj2})
into a direct sum according to:
\begin{equation}
\label{Eq:Dj'D(L-L')/2}
\hat{D}^{j'}({\bf \Omega}) \otimes \hat{D}^{(L-L')/2} ({\bf\Omega})
=
\bigoplus_{j=|j'-(L-L')/2|}^{j'+(L-L')/2} \hat{D}^{j} ({\bf\Omega})
\end{equation}
Inserting Eq.~(\ref{Eq:Dj'D(L-L')/2}) into Eq.~(\ref{Eq:LostTheTrack}) gives:
\begin{equation}
\label{Eq:UOmegaHNLasN-1}
\left. [\hat{U}({\bf\Omega})]^{\otimes N} \right|_{{\cal H}_{NL}}
=
\bigoplus_{L'=0}^{L}
\bigoplus_{j'=L'/2-\lfloor L'/2 \rfloor}^{L'/2}
\bigoplus_{j=|j'-(L-L')/2|}^{j'+(L-L')/2}
\hat{\mathbbm{1}}_{K^{j'}_{N-1,L'}}
\otimes
\hat{D}^{j} ({\bf\Omega})
\end{equation}
Comparing Eq.~(\ref{Eq:UOmegaHNLExpansion}) with the above expression allows us to relate the multiplicities $K^{j}_{NL}$, describing decoherence-free subsystems for $N$ uses of the channel, to the coefficients $K^{j'}_{N-1,L'}$ corresponding to the case when the number of channel uses is reduced by one. In order to find the explicit relation between $K^{j}_{NL}$ and $K^{j'}_{N-1,L'}$ we need to change the order of summations over $L'$, $j'$, and $j$ in Eq.~(\ref{Eq:UOmegaHNLasN-1}) such that the summation over $j$ is the outermost and its limits do not depend on other summation variables. Then the inner summations over $L'$ and $j'$ should yield a formula for $K^{j}_{NL}$. 

\begin{figure}
\begin{center}
\epsfig{file=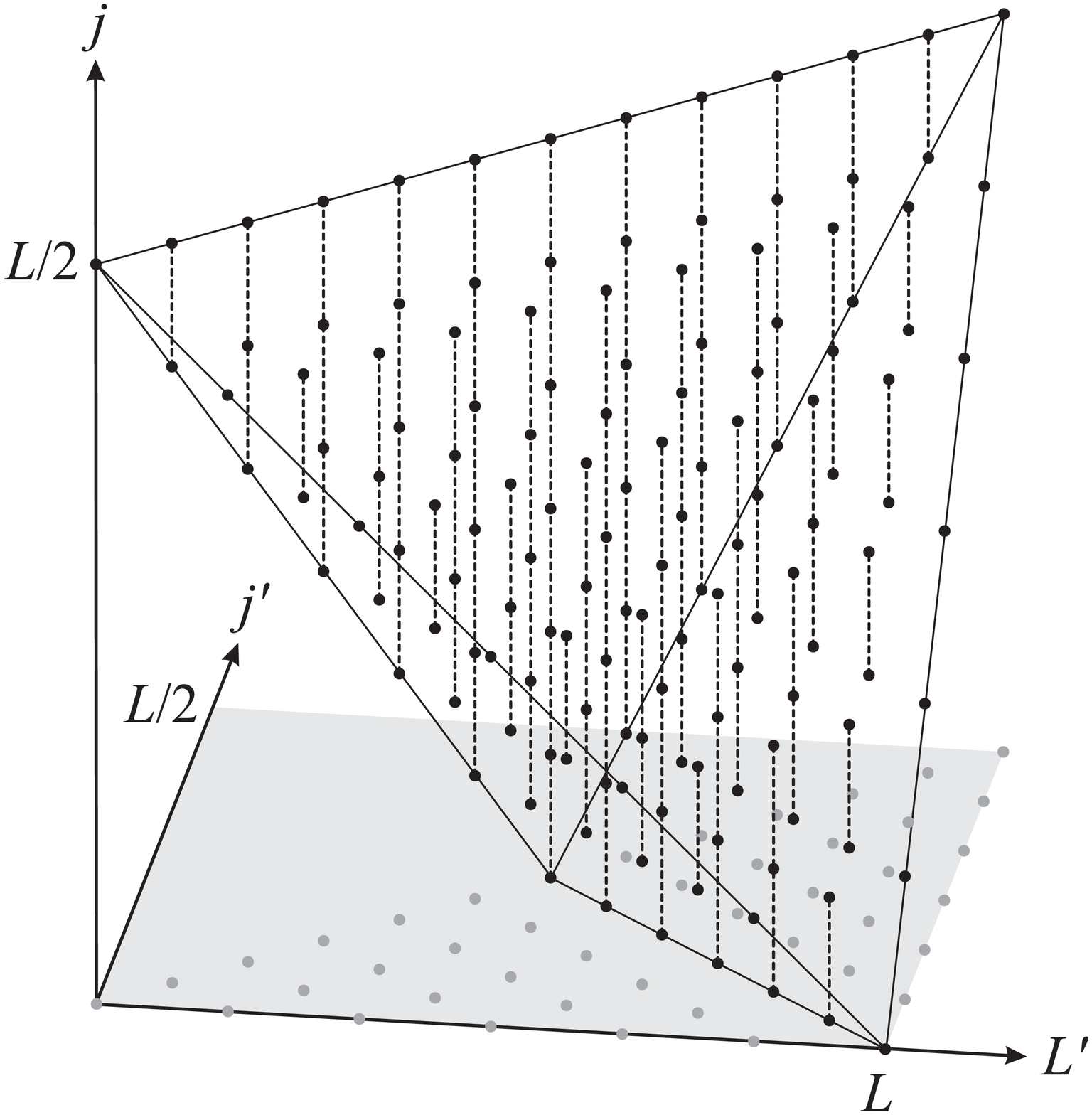,width=10cm}
\end{center}
\caption{The three-dimensional grid for summation variables $(L',j',j)$ defined by
the triple sum in Eq.~(\ref{Eq:UOmegaHNLasN-1}). The allowed values of $L'$ and $j'$
form a triangular grid shown with dark grey points in the $(L',j')$ plane marked
as a light gray square. For every pair of $L'$ and $j'$ the permitted values of
$j$ are shown as black points connected with a dotted line. The entire three-dimensional
grid is contained within a tetrahedron with vertices located at $(L,0,0)$, $(L/2,L/4,0)$,
$(0,0,L/2)$, and $(L,L/2,L/2)$. These points are obtained by calculating intersections of any three out of four planes given by Eqs.~(\ref{Eq:Maxj})--(\ref{Eq:Minj2}) and $j'=L'/2$.}
\label{Fig:3DGrid}
\end{figure}

The triple summation given in Eq.~(\ref{Eq:UOmegaHNLasN-1}) can be visualized by plotting
in a three-dimensional space the set of points $(L',j',j)$ that are defined by the summation limits in Eq.~(\ref{Eq:UOmegaHNLasN-1}). The resulting grid is depicted in Fig.~\ref{Fig:3DGrid}. First, we note that the conditions $L'=0,1,\ldots,L$ and
$j'= L'/2-\lfloor L'/2 \rfloor, \ldots, L'/2$ define a triangular grid for the pairs
$(L',j')$. This two-dimensional grid is shown with the help of dark grey points in 
Fig.~\ref{Fig:3DGrid}. For every pair $(L',j')$ we have an allowed range of $j$s defined
by the limits of the third sum in Eq.~(\ref{Eq:UOmegaHNLasN-1}). The upper limit for $j$ is always given by the plane specified by the equation
\begin{equation}
\label{Eq:Maxj}
j = j'+(L-L')/2.
\end{equation}
The lower limit for $j$ is specified by one of the two conditions:
\begin{equation}
\label{Eq:Minj1}
j=(L-L')/2-j'
\end{equation}
or
\begin{equation}
\label{Eq:Minj2}
j=j'-(L-L')/2
\end{equation}
whichever gives a higher value of $j$. It is easy to verify that Eq.~(\ref{Eq:Minj1}) is relevant when $j'\le (L-L')/2$, whereas Eq.~(\ref{Eq:Minj2}) gives the lower summation limit if $j'\ge (L-L')/2$. The three planes defined by Eqs.~(\ref{Eq:Maxj})--(\ref{Eq:Minj2}) together with the fourth vertical plane specified by
the condition $j'=L'/2$ form a tetrahedron which encloses the entire three-dimensional summation grid.

\begin{figure}[t]
\begin{center}
\epsfig{file=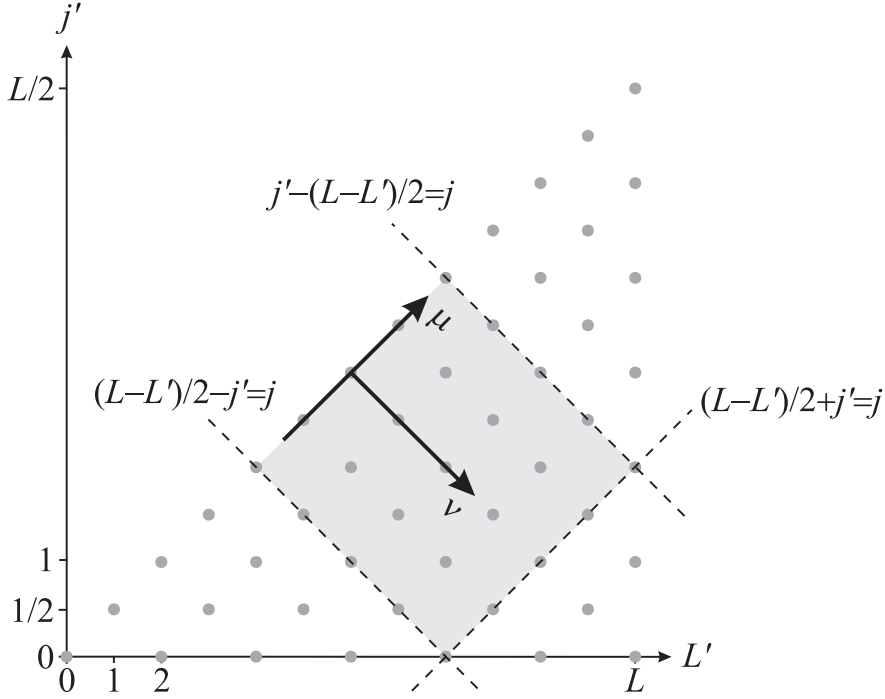,width=12cm}
\end{center}
\caption{The points of the three-dimensional grid $(L',j',j)$
that lie in the horizontal plane for
a specified $j$ have the first two coordinates $(L',j')$ specified by the set
of inequalities $L'/2+j'\ge L/2-j$, $L'/2+j' \le L/2+j$, and $L'/2-j' \le L/2-j$.
These three inequalities, together with the condition $L'/2-j' \ge 0$, define
a rectangular region, which can be conveniently parameterized with the help of
two new integer variables $\mu$ and $\nu$.}
\label{Fig:Square}
\end{figure}

The structure depicted in Fig.~\ref{Fig:3DGrid} provides us with guidance on how to invert the order of summations in Eq.~(\ref{Eq:UOmegaHNLasN-1}), in order to make the summation limits of $j$ independent of other variables. Obviously, the summation over $j$ will run from $L/2-\lfloor L/2 \rfloor$ to $L/2$ in integral steps. In order to find the limits for $L'$ and $j'$ for a fixed $j$, we need to consider an intersection of the tetrahedron depicted in Fig.~\ref{Fig:3DGrid} with a horizontal plane corresponding to that value of $j$. This procedure is illustrated in Fig.~\ref{Fig:Square}. The intersection of the plane of constant $j$ with the planes specified in Eqs.~(\ref{Eq:Maxj})--(\ref{Eq:Minj2}) gives three linear constraints on the values of $L'$ and $j'$ shown in Fig.~\ref{Fig:Square} with dashed lines. These constraints, together with $j'=L'/2$, define a rectangular region which, as it is easy to see, lies entirely within the triangular grid of pairs $(L',j')$. Because of the geometry of this region, it is convenient to define two new variables:
\begin{eqnarray}
\mu & = & L'/2+j', \\
\nu & = & L'/2-j'.
\end{eqnarray}
The limits for $\mu$ are given by Eqs.~(\ref{Eq:Minj1}) and (\ref{Eq:Minj2}) and have the explicit form $L/2-j \le \mu \le L/2+j$, whereas the values of $\nu$ are bounded by $j'=L'/2$ and Eq.~(\ref{Eq:Maxj}) which combined together give $0 \le \nu \le L/2-j$. 
Both the variables $\mu$ and $\nu$ increase in unit steps, which follows from the form of the grid for the $(L',j')$ variables. Consequently, the summation over $L'$ and $j'$ for a specified $j$ can be given by a double sum over $\mu=L/2-j, L/2-j+1, \ldots , L/2+j$ and
$\nu=0,1,\ldots,L/2-j$ with the old variables expressed as $L'=\mu+\nu$ and $j'=(\mu-\nu)/2$.

Using the new summation variables $\mu$ and $\nu$ gives us the following decomposition
for $\left. [\hat{U}({\bf\Omega})]^{\otimes N} \right|_{{\cal H}_{NL}}$
with the inverted integration order:
\begin{equation}
\label{Eq:Umunu}
\left. [\hat{U}({\bf\Omega})]^{\otimes N} \right|_{{\cal H}_{NL}}
=
\bigoplus_{j=L/2-\lfloor L/2 \rfloor}^{L/2}
\bigoplus_{\mu=L/2-j}^{L/2+j}
\bigoplus_{\nu=0}^{L/2-j}
\hat{\mathbbm{1}}_{K^{(\mu-\nu)/2}_{N-1,\mu+\nu}}
\otimes
\hat{D}^{j} ({\bf\Omega})
\end{equation}
The two inner sums yield of course the identity operator acting in a larger space, defined
by:
\begin{equation}
\bigoplus_{\mu=L/2-j}^{L/2+j}
\bigoplus_{\nu=0}^{L/2-j}
\hat{\mathbbm{1}}_{K^{(\mu-\nu)/2}_{N-1,\mu+\nu}}
=
\hat{\mathbbm{1}}_{\sum_{\mu=L/2-j}^{L/2+j}
\sum_{\nu=0}^{L/2-j}
K^{(\mu-\nu)/2}_{N-1,\mu+\nu}}.
\end{equation}
Inserting this equation into Eq.~(\ref{Eq:Umunu}) and comparing the resulting expression with Eq.~(\ref{Eq:UOmegaHNLExpansion}) gives a recursion formula for $K^{j}_{NL}$ in the form:
\begin{equation}
\label{Eq:KRecursion}
K^{j}_{NL} = 
\sum_{\mu=L/2-j}^{L/2+j}
\sum_{\nu=0}^{L/2-j}
K^{(\mu-\nu)/2}_{N-1,\mu+\nu}
\end{equation}
which is the central result of this paper. In order to complete the recipe for calculating the multiplicities $K^{j}_{NL}$, we need to specify their values for $N=1$. This task however is straightforward. Let us recall that for $N=1$ we have ${\cal H}_{N=1,L}={\cal H}^{(L)}$
where the right-hand side has been defined in Eq.~(\ref{Eq:H(l)=Span}), and that for a polarization transformation ${\bf\Omega} \in \mbox{SU(2)}$ Eq.~(\ref{Eq:UOmega}) implies that $\left.\hat{U}({\bf\Omega})\right|_{{\cal H}_{N=1,L}} = \hat{D}^{L/2}({\bf\Omega})$.
This means that for $N=1$:
\begin{equation}
K^{j}_{N=1,L} = \delta_{L,2j}
\end{equation}
which provides the initial condition to calculate $K^{j}_{NL}$ for an arbitrary $N$. 

\begin{figure}[t]
\begin{center}
\epsfig{file=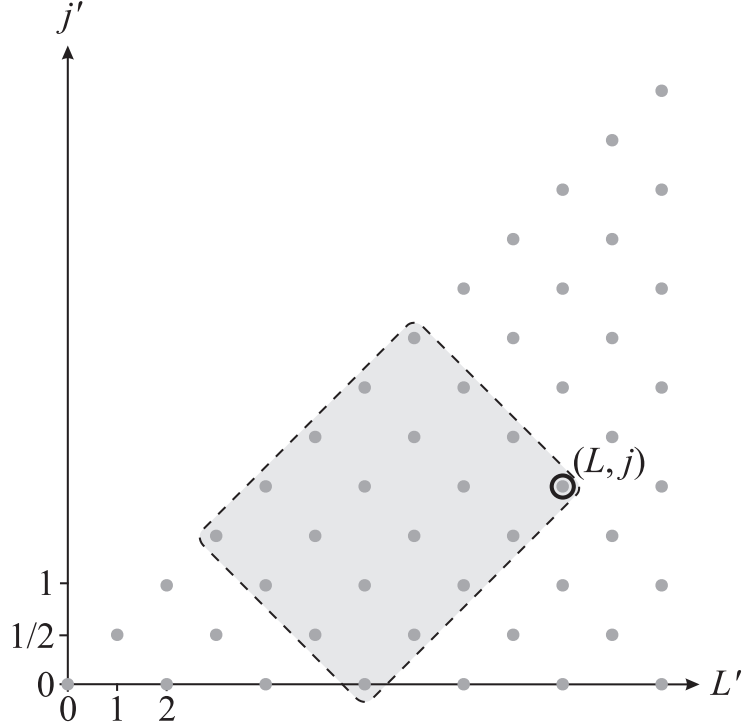,width=10cm}
\end{center}
\caption{Graphic representation of the recursion formula derived in Eq.~(\ref{Eq:KRecursion}). The multiplicity $K^{j}_{NL}$ is given by a sum of the multiplicities $K^{j'}_{N-1,L'}$ for all the pairs of the parameters $(L',j')$ lying inside a rectangle marked as a light
grey region, with vertices at $(L,j)$, $(L-2j,0)$, $(L/2-j,L/4-j/2)$, and $(L/2+j,L/4+j/2)$.}
\label{Fig:Recursion}
\end{figure}

The recursion formula derived in Eq.~(\ref{Eq:KRecursion}) can be illustrated with a diagram shown in Fig.~\ref{Fig:Recursion} which provides a mnemonic recipe for carrying out calculations. In order to find $K^{j}_{NL}$ we need to add $K^{j'}_{N-1,L'}$ for all pairs of $(L',j')$ that lie within a rectangle which is rotated by $45^{\circ}$ with respect to the axes of the coordinate system, and has two vertices located at $(L,j)$ and $(L-2j,0)$, with the remaining two located on the line $j'=L'/2$. Fig.~\ref{Fig:Recursion} shows how to obtain this rectangle by starting from the point $(L,j)$ and going in two orthogonal directions that form $45^{\circ}$ with the axes, until reaching respectively the limits of $j'=L'/2$ and $j'=0$.

\section{Conclusions}
\label{Sec:Conclusions}

We have derived a recursion formula for the dimensions $K^{j}_{NL}$ of decoherence-free subsystems in a bosonic channel experiencing collective depolarization described by the U(2) group. This model is relevant in quantum communication over single-mode optical fibers, for which collective depolarization is one of the dominant decoherence mechanisms. Although we have not been able to solve the recursion formula and obtain a closed analytical formula for $K^{j}_{NL}$, it can be easily implemented in numerical calculations.

The depolarization model considered here assumed that the phase relations between the consecutive uses of the channel are fixed. This requirement cannot be fulfilled if the communicating parties do not share a common phase reference to prepare and detect states. Then the phase factor $e^{-i\alpha}$ varies between the uses of the channel, and only the SU(2) transformation ${\bf\Omega}'$ remains constant. In order to implement decoherence-free encoding in such a scenario, the same fixed number of $l$ excitations must be transmitted in a sequence of channel uses.
The structure of decoherence-free subsystems for $N$ uses of the channel is then given by a direct-sum decomposition of $[\hat{D}^{l/2}({\bf\Omega}')]^{\otimes N}$, discussed in \cite{HolbKribXXX04}. Alternatively, it is possible to devise schemes in which phases are
self-referenced by employing multiport interferometers for state preparation
and detection. Examples of such schemes have been described in \cite{WaltAbourPRL03,BoilLaflPRL04}. Mathematically, this approach consists in introducing
phase dependencies between multiple uses of the channel, which then give rise to non-trivial decoherence-free subsystems.

{\em Acknowledgements.} This work was supported by the UK Engineering and Physical Sciences
Research Council and by Polish Committee for Scientific Research, Project No. PBZ KBN 043/P03/2001.

\end{document}